\documentclass{IEEEcsmag_PREPRINT}

\usepackage[colorlinks,urlcolor=blue,linkcolor=blue,citecolor=blue]{hyperref}

\usepackage{upmath}

\usepackage{balance}

\jvol{}
\jnum{}
\paper{}
\jmonth{}
\jname{}
\pubyear{}

\begin{document}

\sptitle{}
\editor{E}

\title{Information Visualization for Effective Altruism}

\author{\vspace{-3mm}Pierre Dragicevic}
\affil{Univ. Bordeaux, CNRS, Inria, LaBRI, France\vspace{-5mm}}

\markboth{Information Visualization for Effective Altruism}{}

\newcommand{\permission}{}

\begin{abstract}
Effective altruism is a movement whose goal it to use evidence and reason to figure out how to benefit others as much as possible. This movement is becoming influential, but effective altruists still lack tools to help them understand complex humanitarian trade-offs and make good decisions based on data. Visualization -- the study of computer-supported, visual representations of data meant to support understanding, communication, and decision making -- can help alleviate this issue. Conversely, effective altruism provides a powerful thinking framework for visualization research that focuses on humanitarian applications.
\end{abstract}

\maketitle
\newcommand{\shortquote}[1]{``\emph{#1}''}


\textbf{2024 update}: in the light of revelations about serious problems in some branches of the effective altruism movement, I no longer support the movement as a whole 
(\href{https://dragice.fr/ea.html}{more info}).

\begin{figure}
\centerline{\includegraphics[width=18.5pc]{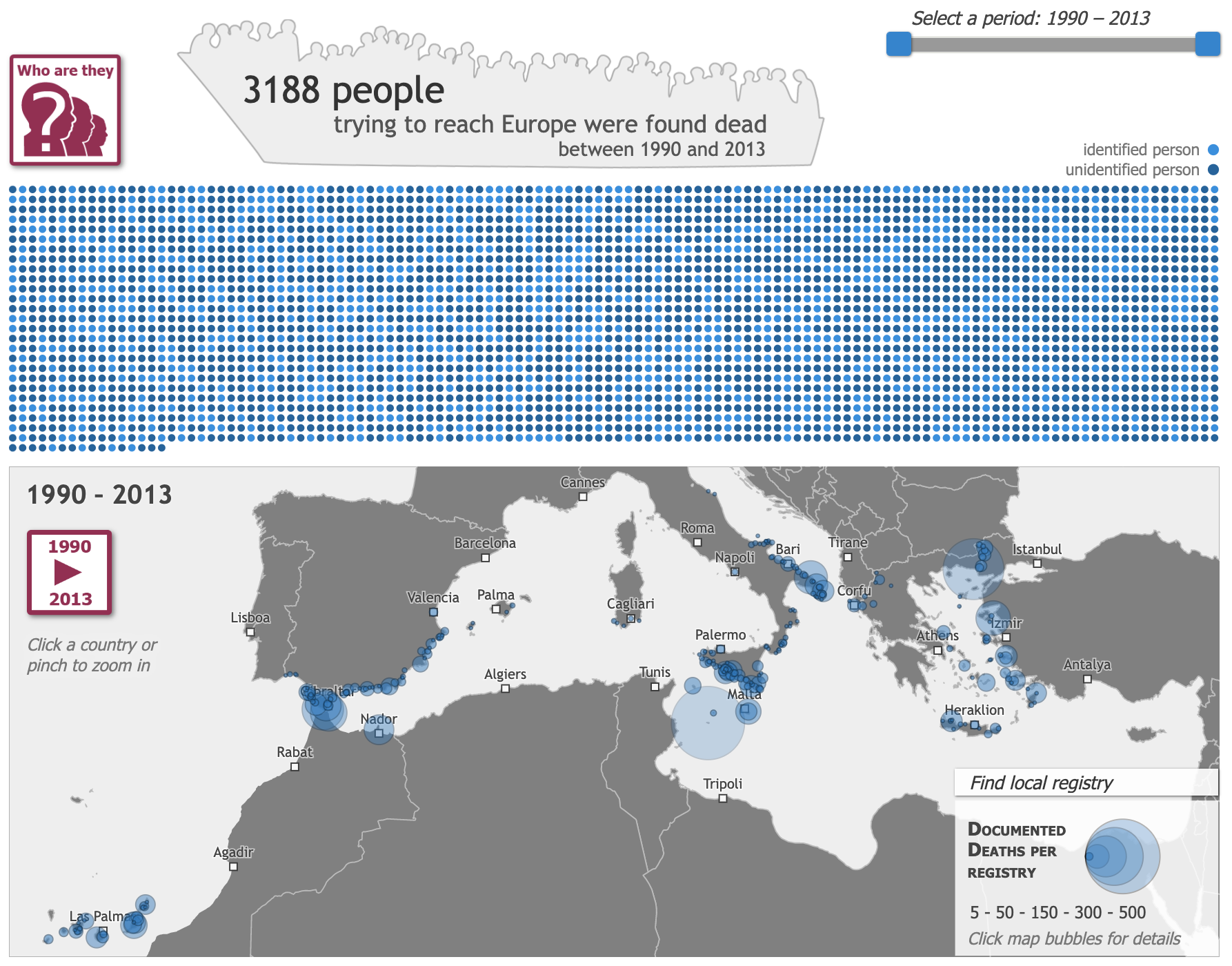}}
\caption{Interactive web visualization showing deaths at the borders of Southern Europe between 1990 and 2013. Each small dot represents one dead migrant documented by local authorities. Source \url{http://www.borderdeaths.org/}. \permission}
\label{fig:borderdeaths}
\end{figure}

\chapterinitial{A lot} has been written on how data visualizations can be useful for a variety of tasks~\cite{fekete2008value}, but only recently have researchers started to consider how they can be used to promote human welfare. In particular, visualization researchers have started to look at how data journalists use data visualizations to raise the public awareness about humanitarian issues (see \autoref{fig:borderdeaths}). They coined the term \textit{anthropographics} to refer to \shortquote{visualizations that represent data about people in a way that is intended to promote prosocial feelings or prosocial behavior} \cite{morais2022showing}. More recently, the term \textit{humanitarian visualization} was introduced to refer to \shortquote{data visualizations or infographics designed to promote human welfare} \cite{dragicevic2022towards}.

Despite the great potential of humanitarian visualization as a practice and a research area, its ability to make progress and have a positive impact is limited by the way problems are generally framed, and by the types of solutions and metrics of success typically considered. Suppose a newspaper publishes a striking infographic about the number of blind persons in the U.S. who need a guide dog, resulting in lots of people donating to a charity that provides guide dogs. Most practitioners and researchers would certainly consider this infographic a success, and a benefit to society. This is clearly the case if the counterfactual is a world where the same donors keep the money for themselves. However, the situation is less clear if the counterfactual is a world where the donors choose to give their money to a charity that provides even more benefits. In fact, training a guide dog in the U.S costs \$40,000, and the same amount of money can be used to cure more than 2,000 people in Africa of blindness by paying for surgeries to reverse the effects of trachoma \cite{ord2013moral}. How money should be allocated in this case is a matter of personal judgment, but a lot can be gained by helping people -- both the general public and decision makers -- allocate their limited resources in an informed manner, with sufficient knowledge about the range of options they have at their disposal, and their expected impact.

\begin{figure}
\centerline{\includegraphics[width=18.5pc]{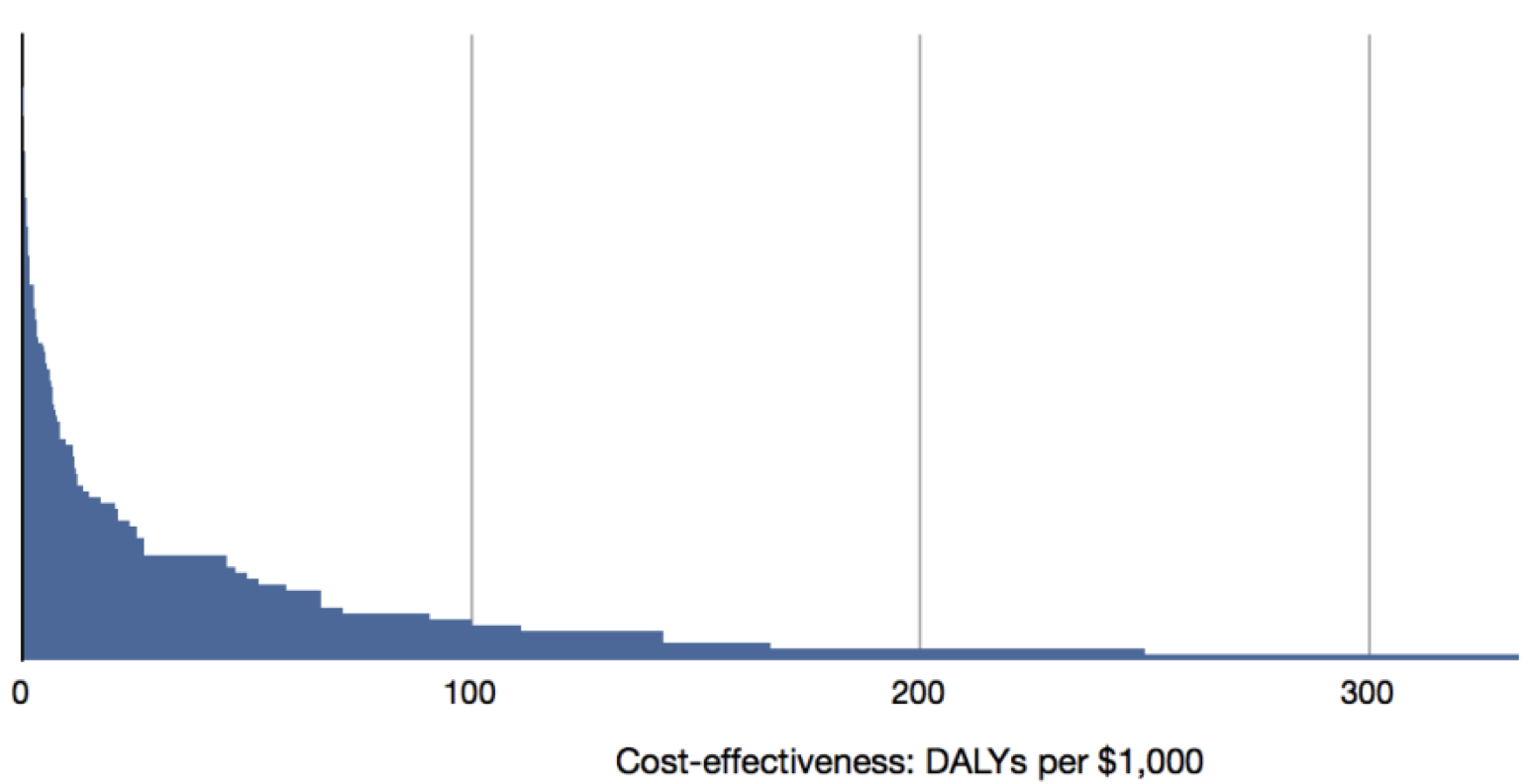}}
\caption{Distribution of the cost-effectiveness of 100+ health interventions, expressed in disability-adjusted life years (DALYs) per \$1,000. Source \cite{ord2013moral}. \permission}
\label{fig:effectiveness-distribution}
\end{figure}

Impact would not be a central concern if charity programs and interventions were roughly comparable in terms of their human impact, but this is very far from being the case. For example, a global health study has estimated the effectiveness of more than 100 health interventions in terms of disability-adjusted life years (DALYs) saved per \$1,000, finding a remarkably wide and skewed distribution (see \autoref{fig:effectiveness-distribution}). Effectiveness ranges from 0.02 to 300 depending on the intervention (a factor of 15,000), with a  median of 5, suggesting that the vast majority of health interventions are far less effective than the most effective ones. Thus, moving money from the many ineffective interventions to the most effective ones is likely to be helping people considerably more than donating even a lot of money to a random intervention.

Yet donating money to a random intervention is precisely the kind of task currently supported by most humanitarian visualizations, due to their focus on \textit{case-by-case persuasion}: a cause is pre-identified, and the goal is to raise people's concerns about that cause as much as possible. Similarly, in research, the focus is on finding the design strategies that are the most effective independently from the cause. However, strategies that help make a message persuasive are not necessarily the ones that promote the best decisions. As an example, a lot has been written on the power of storytelling in visualization \cite{kosara2013storytelling}; But as Neil Halloran, the designer of the celebrated data-driven documentary \textit{The Fallen of WWII} has pointed out, \shortquote{you can tell a story about a crisis of any size, and tell a compelling story} \cite{halloran2017emotional}, implying that stories do not necessarily help people to think rationally about the extent of human suffering. Visualization researchers could greatly benefit from new thinking frameworks to help them move beyond case-by-case persuasion and reason about how to best use visualization to alleviate human suffering on a global scale. Effective altruism provides such a framework.

\section{What is Effective Altruism?}

The term \textit{effective altruism} (often abbreviated as \textit{EA}) was coined in 2011 at Oxford University, by a small group of academic philosophers and individuals involved in charity and philanthropy organizations \cite{macaskill2019definition}. In 2016, the head of this group, William MacAskill, worked with many leaders involved in the EA community to write a definition that has been widely endorsed by the community:

\vspace{1mm}
\begin{quote}
    \textit{Effective altruism is about using evidence and reason to figure out how to benefit others as much as possible, and taking action on that basis.} \cite{macaskill2019definition}
\end{quote}
\vspace{1mm}

In 2018, using again input from many EA leaders, MacAskill proposed a more precise definition to be used in academic discussions:

\vspace{1mm}
\begin{quote}
\textit{Effective altruism is: (i) the use of evidence and careful reasoning to work out how to maximize the good with a given unit of resources, tentatively understanding ‘the good’ in impartial welfarist terms, and (ii) the use of the findings from (i) to try to improve the world.} \cite{macaskill2019definition}
\end{quote}
\vspace{1mm}

This definition highlights the double aspect of EA as \textit{(i)} an intellectual project (a research field) and \textit{(ii)} a practical project (a social movement). The definition is non-normative: it does not say how people should behave (e.g., that we should make personal sacrifices to help others). \textit{Welfarist} means that views that assign intrinsic value to other things than well-being (e.g., biodiversity, art, or knowledge) are excluded, while \textit{impartial} means that views that do not weigh people equally (e.g., prioritizing nationals over foreigners) are also excluded. \textit{Tentative} means that the impartial welfarist view is a working assumption that can be debated and refined within EA. For example, while animal welfare is a central concern for many EA proponents, how much moral weight should they be given compared to humans remains unclear.

These broadly-accepted definitions are very helpful when discussing the merits and weaknesses of EA, because many criticisms of EA arise from people using their own interpretation of what it is. This leads to common misconceptions, such as: EA is just applied utilitarianism, it is only about fighting poverty, it is only about donations or earning to give, and it ignores systemic change (for discussions see \cite{macaskill2019definition}).

\section{How can EA Inform Visualization}

Because people have different moral intuitions, not all researchers working on -- or considering working on -- humanitarian visualization will find the EA philosophy compelling enough to embrace it. But for those who do, EA can provide a clear thinking framework in an area that has been lacking one. Indeed, many discussions so far have focused on how to design visualizations that elicit empathy, often ignoring that empathy does not necessarily promote helping behavior \cite{morais2022showing}. Even when a visualization does cause people to act, their actions can have a negligible, null, or possibly even negative impact on global human welfare. EA provides clear grounds to think about research goals and metrics of success.

The EA lens can also help researchers think out of the box and broaden the scope of humanitarian visualization research by identifying new types of solutions and approaches. In particular, some visualizations may not promote prosocial feelings or behavior -- and thus might not be considered conventional humanitarian visualizations -- but may still promote welfare. For example, a visualization that helps a charity director effectively allocate money across different health programs does not promote prosocial feelings or behavior (since all the money will be used to help people no matter what), but it can tremendously increase human welfare.


\begin{figure*}
\centerline{\includegraphics[width=0.9\textwidth]{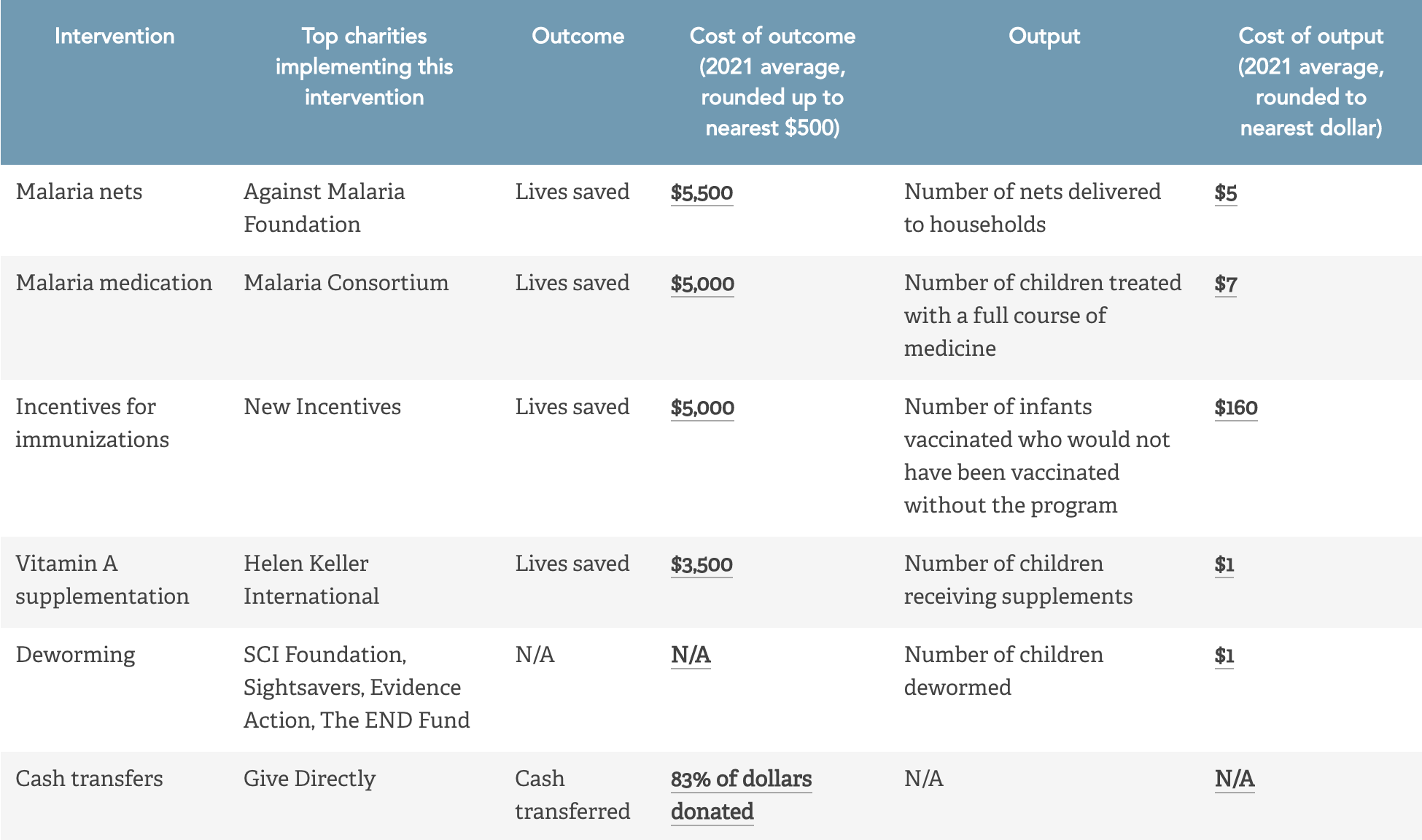}}
\caption{Table showing impact metrics for six charities identified as among the most effective by GiveWell in 2021. Source \url{https://www.givewell.org/cost-to-save-a-life}. \permission}
\label{fig:givewell-table}
\end{figure*}

EA is a thinking framework but it is also a community. This community is full of people who are deeply knowledgeable about humanitarian issues or have been extensively involved in humanitarian actions, and thus visualization researchers could learn a lot by connecting with them. In addition, the EA community has unique needs that visualization could help address. For example, several EA organizations do research on the effectiveness of different charities and charity programs, in order to guide potential donors. GiveWell is a known example: it maintains a list of top effective charities, primarily based on the cost of life saved (see \autoref{fig:givewell-table}). GiveWell shares a range of spreadsheets with data and calculations to explain how it arrived at its estimates. All such initiatives generate lots of useful data, but the amount of information can rapidly become overwhelming for potential donors. And yet, data is currently largely communicated through numbers and text, and very rarely through visualizations. Perhaps visualization could also be used by the EA communicators to better explain its general principles to naive audiences, and by EA researchers to help them analyze the effectiveness of different charity programs.

\section{Using Visualization and Psychology to Support EA}

People share many misconceptions and biases preventing them from helping effectively -- for example, geographical and cultural proximity often greatly affect how much people feel like helping \cite{caviola2021psychology}. Researchers studying humanitarian visualization can take inspiration from recent work on judgment and decision making with visualizations \cite{dimara2018task}, and apply findings and methods from psychology to study how visualizations interact with cognitive biases, and whether visualizations can help alleviate those biases. Unfortunately, much like visualization research, psychology research has mostly focused on how to make people donate more, rather than more effectively. However, Lucius Caviola and colleagues \cite{caviola2021psychology} have recently done a tremendous job at reframing past findings through an EA lens, leading them to identify major psychological obstacles to effectiveness, which fall in two categories:

\textit{1. Motivational obstacles.} People think that whether and how to help is largely a matter of personal preference; they give based on how much they feel emotionally connected to the issue (e.g., they feel more strongly about diseases that are common in their country or have affected their loved ones); they dislike prioritizing some causes over others; they view people who try to donate rationally more negatively than those who donate based on empathy.

\textit{2. Epistemic obstacles.} People think that charity overhead is wasteful, or find funding overhead unsatisfying; they think that effectiveness cannot be quantified; they do not think clearly about probabilities; they are not aware that charities differ greatly in their effectiveness; they don't know which charities are the most effective.

Caviola and colleagues also identified four types of strategies to increase effective giving: information, choice architectures and incentives, philosophical reasoning,  and norm changes.

\textit{Information} addresses epistemic obstacles through education. As I mentioned before, one of the areas where visualization can help is by conveying rich quantitative facts about charity effectiveness in a way that is easy to process. Visualization could also be used to argue for lesser-known EA causes such as wild animal suffering and global catastrophic risks, by conveying data about how serious, neglected and tractable these causes are. Finally, visualization could also help dispel misconceptions, for example by showing data about how charity overhead is employed, together with simulations illustrating how cutting overhead would likely yield less positive outcomes. Information is the type of strategy where the possible benefits of visualizations are the most evident, and where a lot can be done in collaboration with the EA community.

\textit{Choice architectures and incentives} address motivational obstacles by nudging (e.g., using effective charity programs as default options) and incentivization (e.g., using donation matching or tax deductions targeted to effective charity programs). Here, possible roles for visualization are less immediately evident, but this type of strategy can potentially lead to the most interesting innovations and contributions to knowledge. In particular, it could be interesting to study which nudging techniques can translate to visualizations. For example, a well-documented bias is the decrease of people's concern for individual victims as the number of victims increases: a tragedy that affects one million people typically does not generate 100 times more concern or donations than a tragedy that affects a thousand people \cite{caviola2021psychology}. However, this effect is less pronounced when donors evaluate all options at the same time -- which could mean seeing data about multiple tragedies visualized side-by-side -- than if they evaluate the options sequentially.

The last two categories listed by Caviola and colleagues, \textit{philosophical reasoning} (exposing people to philosophical arguments) and \textit{norm changes} (pushing for a change of moral standards) are important but probably less directly relevant to visualization.

\section{Conveying Personal Experiences with Quantitative Facts}

Again, a major way in which visualization can support EA is by helping people compare charity programs. To take a trivial example, an EA website could include as an overview of its top programs a bar chart of the number of lives saved per unit of donation for each program.

\vspace{3mm}
\noindent\fbox{\parbox{6.8cm}{\small{
Sources for the figures mentioned in this section:
\begin{itemize}
    \item Against Malaria Foundation (2022), Why nets? \url{https://www.againstmalaria.com/WhyNets.aspx}
    \item F. Ricci (2021) Social implications of malaria and their relationships with poverty. \textit{Med. j. of hematology and infectious diseases.}
    \item World Health Organization (2022) Vitamin A deficiency \url{https://www.who.int/data/nutrition/nlis/info/vitamin-a-deficiency}
    \item Stephen Clare (2020) Homelessness in the US and UK Executive Summary \url{https://www.founderspledge.com/stories/homelessness-in-the-us-and-uk-executive-summary}
    \item John Halstead (2019) Founders Pledge -- Mental Health Executive Summary \url{https://founderspledge.com/stories/mental-health-report-summary}\vspace{-2mm}
\end{itemize}
}}}
\vspace{2mm}

However, this is only a minimalist example, and important visualization design challenges arise when a variety of outcomes need to be visualized and compared. For example, about 600 mosquito nets prevent the death of a child, but they also prevent 500 to 1,000 cases of malaria. This is an enormous benefit in and of itself, as malaria is a crippling disease with flu-like symptoms that can periodically return, can be highly disruptive for the life of households, and can leave children disabled. Similarly, GiveWell lists a charity that saves lives by giving vitamin A supplements to children. But even when it is not fatal, vitamin A deficiency causes a range of terrible problems such as repetitive infections and blindness. GiveWell sometimes goes beyond lives saved and considers charities expected to impact the recipient's lifetime earnings (treatments for parasitic worm infections) or their overall quality of life (cash transfers for extreme poverty). Another effective altruism website lists a charity that can use about \$20,000 to prevent a year of homelessness in the US or UK, and another one that can use \$200--\$300 to prevent the equivalent of one year of severe major depressive disorder for a woman in Uganda. It is very hard to imagine how to visualize those widely different types of outcomes in a way that supports informed, effective-altruist decisions.

Ideally, a major donation or funds allocation decision should be based both on quantitative facts (e.g., the number of people affected, the cost of interventions) and a deep understanding of people's subjective experiences with and without the interventions, especially concerning the degree of physical and psychological suffering involved. However, it is hard for a person who has never contracted malaria or never had a vitamin A deficiency to have a reliable intuition of what those experiences entail. This is where stories -- in the form of text, images, graphic novels, movies or video games -- could play an important role by helping people understand subjective experiences on a visceral level. I have previously emphasized the limits of storytelling for EA purposes, but certain ways of combining stories with data may be very effective at supporting EA.

One potentially effective strategy could be to \textit{(i)} use stories to give a qualitative understanding of the personal experiences involved in a human tragedy, and \textit{(ii)} use data to give a quantitative understanding of the extent of the tragedy. It seems important that both elements are provided in order to support EA decisions. In particular, stories of personal tragedies provide a proof of existence but can give a distorted vision of reality in the presence of selection bias: news media, for example, often select atypical stories based on their shock value. But if personal stories are complemented with clear data about how representative they are, viewers will get a more accurate appreciation of the extent of the problems and of the magnitude of the human suffering involved.

\begin{figure}
\centerline{
\includegraphics[height=6.6pc, trim={4mm 0 12mm 0}, clip]{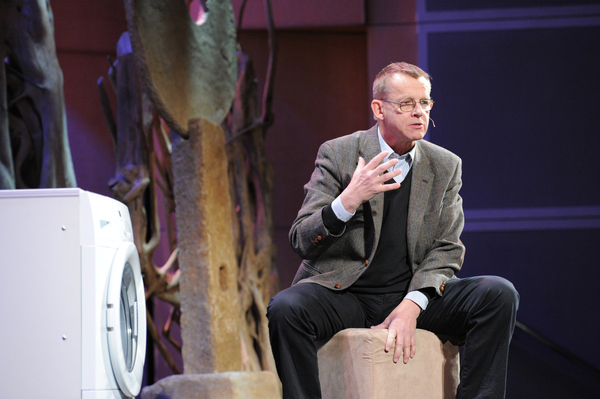}
\includegraphics[height=6.6pc]{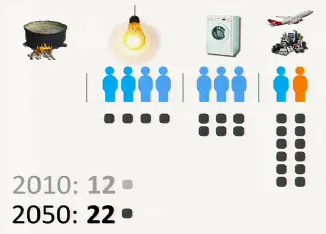}
}
\caption{Images from Hans Rosling's talk at TEDWomen 2010. Sources \url{https://www.ted.com/talks/hans_rosling_the_magic_washing_machine} and Flickr. Left photo by James Duncan Davidson / TED. Used with permission.}
\label{fig:rosling}
\end{figure}

It will likely be a major research challenge to find out more concretely how to effectively combine stories with quantitative data. There are at least two possible approaches: in a \textit{data-then-story} approach, people would view statistical data about tragedies or social issues, and then zoom into individuals to see their personal stories, either real or hypothetical. The choice of individuals may be decided by the viewer following a details-on-demand approach, or it may follow a random sampling scheme. Meanwhile, in a \textit{story-then-data} approach, people would first see one or several typical stories (for example, the daily life of someone with disease A or disease B), and would then be able to explore statistical data (for example, the prevalence of those two diseases, and how they could be reduced with different interventions). An example of story-then-data approach is Hans Rosling's talk \textit{The Magic Washing Machine} (\autoref{fig:rosling}): he first tells a story that gives a powerful account of how life-changing washing machines are, and then goes through data about how many people in the world have access to them, and how this is likely to change with economic growth.

It is challenging to reconcile the world of numbers with the world of subjective experience, but not impossible -- for example, if an effective altruist judges that having disease A is twice as bad as having disease B, they could conclude that preventing 10,000 cases of disease A is equally desirable as preventing 20,000 cases of disease B.

\section{Emerging Technologies}

In visualization research, there has been a lot of interest in conveying visualizations through other media than computer screens, like physical objects \cite{dragicevic2020data} and mixed reality displays \cite{kraus2021value}. In a previous position paper \cite{dragicevic2022towards}, I discuss the interesting research opportunities offered by such media for the purpose of humanitarian visualization. I summarize them here.

\begin{figure}
\centerline{\includegraphics[width=18.5pc]{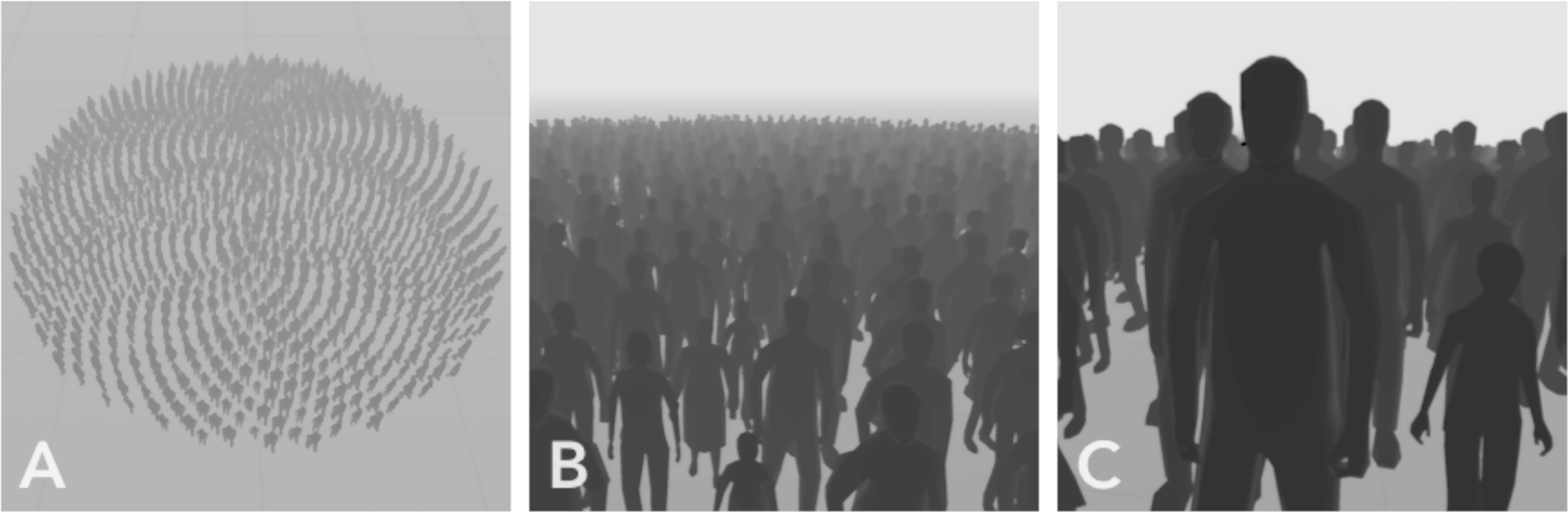}}
\caption{VR visualization of mass shooting data in the US. Source \cite{ivanov2019walk}. \permission}
\label{fig:ivanov}
\end{figure}

\textit{Virtual reality}. By providing a way for viewers (e.g., donors or charity managers) to immerse themselves more fully into personal stories, virtual reality (VR) may help enhance their visceral understanding of human issues and tragedies. VR documentaries already exist that cover topics such as war, migration, and diseases. Such immersive stories could be combined with immersive data visualizations for EA purposes. This idea has started to be explored by Ivanov and colleagues \cite{ivanov2019walk}, who designed a VR visualization of mass shooting casualties in the US (\autoref{fig:ivanov}). Each silhouette represents a person who died from a mass shooting in the US. Viewers can step back to get an overview of the dataset (see panel A in \autoref{fig:ivanov}), or come closer to gather information about individual victims such as their age group or gender, which are encoded by the shape of the silhouette (panels B and C).

The concept from Ivanov et al. is only a starting point, as one could imagine conveying richer qualitative information about each victim like their physical appearance (as some memorials do by showing photo portraits) or elements of their personal stories, which viewers could choose to relive from a first-person perspective. Unfolding or hypothetical humanitarian issues could be conveyed in a similar manner using a combination of data visualizations and immersive video footage (such as already used in VR documentaries) or simulated scenes. VR could also be used to convey the positive outcomes of donations; For example, one could imagine an immersive version of GDLive (\url{https://live.givedirectly.org/}), a website that posts information and updates about recipients of cash transfers.

\textit{Augmented reality.} Augmented reality (AR) can create illusions of objects and people around us, including objects and people that exist remotely. This opens up unprecedented possibilities for bringing the lives of distant suffering people closer to our own, and making humanitarian issues more salient or more memorable. For example, if a person walks in a refugee camp that has been temporarily relocated in their backyard, they may create a mental association and remember the refugees each time they see (or even think about) their backyard. Visualizing data about refugee camps in such a way could thus give a much more lasting impression. In contrast, VR can subjectively transport viewers in distant places, but once the viewers are back, the event is remote again. As with VR, AR could also be used to convey positive outcomes of charitable donations. In the context of a donor/recipient pairing program, future AR technology may even make it possible for a donor to meet a past recipient on the street and chat with them: a long-distance cash transfer may suddenly feel like helping out an acquaintance in a small village. Finally, in the future, effective altruists may be able to use wearable AR devices as commitment devices, e.g., to get regularly reminded of remote tragedies or ways they can redirect unnecessary personal expenses to humanitarian causes.

\begin{figure}
\centerline{\includegraphics[width=18.5pc]{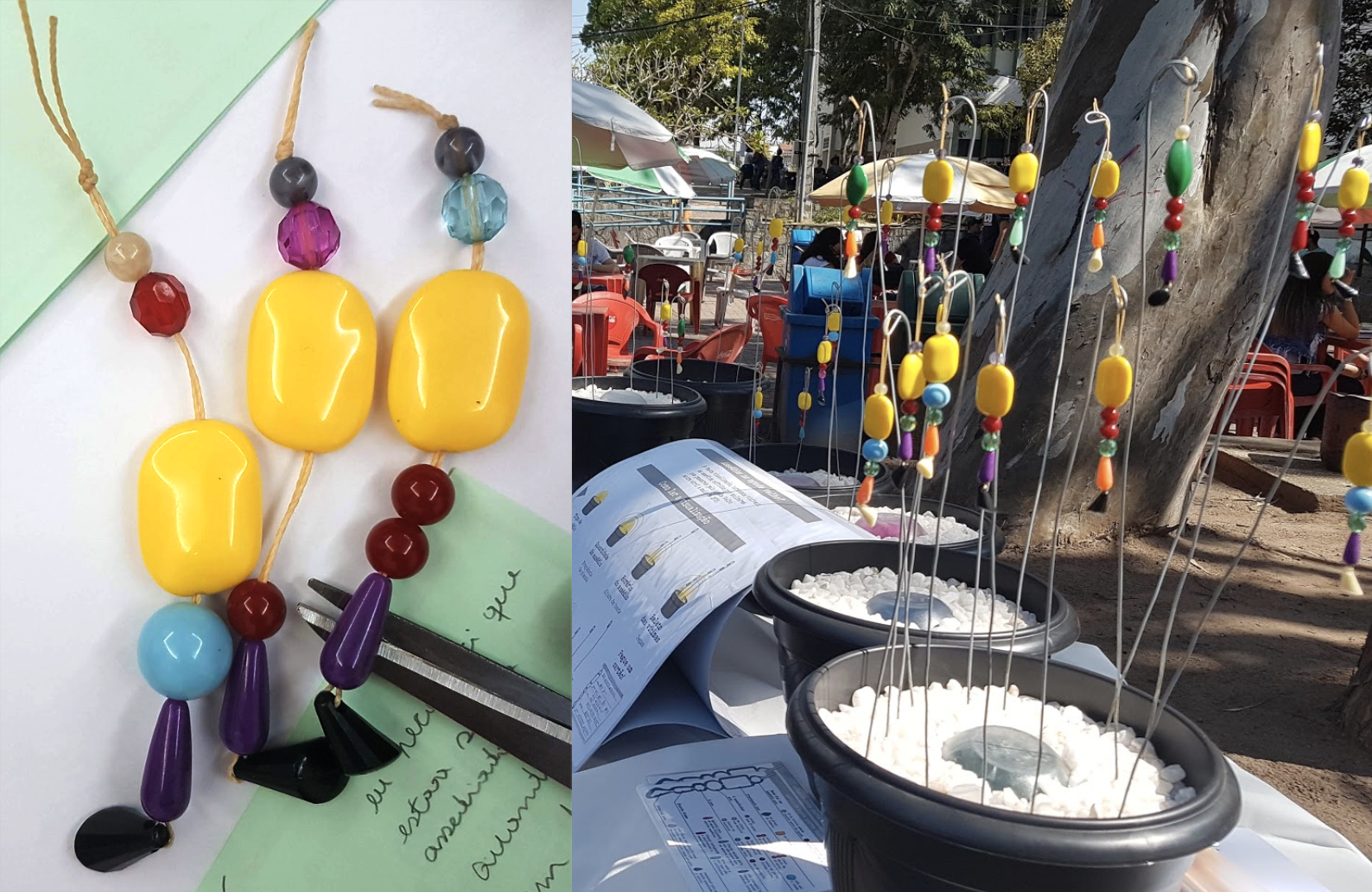}}
\caption{Physical data visualizations of 28 cases of sexual harassment -- each plant conveys data reported by one person. Source \cite{morais2022exploring}. \permission}
\label{fig:plants}
\end{figure}

\textit{Data physicalization}. Data physicalizations are physical entities whose shape or geometry encodes data \cite{dragicevic2020data}. Public spaces already contain physical objects that convey past human tragedies, such as memorials, sculptures and cemeteries. However, few of them focus on current issues and few of them convey quantitative facts, both of which are important for EA purposes. Rare exceptions include data sculptures (artistic data physicalizations) with a focus on humanitarian data, and occasional explorations by visualization researchers like the \textit{Harassment Plants} (\autoref{fig:plants}). Like AR visualizations, physical visualizations can be embedded in our everyday environment. But unlike AR visualizations, they are always present, they can be touched, and do not need special equipment to be seen. On the other hand, AR content can be created and displayed at will.

\textit{Ambient displays}. Ambient displays are displays that \shortquote{present information within a space through subtle changes in light, sound, and movement, which can be processed in the background of awareness} \cite{wisneski1998ambient}. In particular, research projects have explored how ambient displays can support remote intimacy -- for example, the color of a lamp may change according to the affective state of a remote intimate partner captured through a wearable biofeedback device. Similar devices could be used to convey quantitative information about the plight of large populations of distant and anonymous people, such as the number of hospitalizations during a pandemic or the number of war casualties. Such ambient displays could give a continuous impression of the severity of an ongoing humanitarian crisis without having to constantly poll news reports.

\section{CONCLUSION}

Effective altruism offers both a new thinking framework and new questions and problems for visualization research. Yet, it appears that there has been virtually no collaboration so far between EA actors and visualization researchers, perhaps largely due to a lack of mutual awareness between the two communities. But this is changing, as EA is becoming mainstream and highly influential \cite{matthews2022how}. There are many fascinating questions and problems at the intersection of the two areas and unique opportunities for collaboration, so it is time visualization researchers reach out to the EA community and vice versa.

\nocite{*}

\balance 

\bibliographystyle{IEEEtran}
\bibliography{bibliography}

\begin{IEEEbiography}{Pierre Dragicevic} is a permanent Research Scientist at Inria (France) since 2007, studying information visualization and human-computer interaction. He received his PhD from the Université de Nantes in 2004 and was a post-doctoral fellow at the University of Toronto from 2006–2007. His research interests include humanitarian visualization, physical and immersive visualizations, judgment and decision making with visualizations, research transparency and statistical communication, as well as design spaces and conceptual frameworks. Contact him at pierre.dragicevic@inria.fr.
\end{IEEEbiography}

\end{document}